\documentclass{aa}
\bibpunct{(}{)}{;}{a}{}{,} % to follow the A&A style
\pdfoutput=1
\usepackage{graphicx}
\usepackage{txfonts}
\usepackage{mathtools}
\usepackage[colorlinks=true,linkcolor=blue,urlcolor=blue,citecolor=blue]{hyperref}
\makeatletter
\renewcommand*\aa@pageof{, page \thepage{} of \pageref*{LastPage}}
\makeatother

\begin{document} 

\title{Substructures, Resonances and debris streams}
\subtitle{A new constraint on the inner shape of the Galactic dark halo}

\author{Emma Dodd \inst{1}
        \and
        Amina Helmi \inst{1}
        \and
        Helmer H. Koppelman \inst{2}
          }

 \institute{Kapteyn Astronomical Institute, University of Groningen, Landleven 12, 9747 AD Groningen, The Netherlands\\
              \email{dodd@astro.rug.nl}
\and
   School of Natural Sciences, Institute for Advanced Study, Princeton, NJ 08540, USA
}

 \date{Received xxxx; accepted yyyy}

  \abstract
  % context heading
   {The local stellar halo of the Milky Way contains the debris from several past accretion events.}
  % aims heading 
   {Here we study in detail the structure and properties of nearby debris associated with the Helmi streams, originally identified as an overdensity in integrals of motion space.}
  % methods heading 
   {We use 6D phase-space information from \textit{Gaia} EDR3 combined with spectroscopic surveys, and we analyse the orbits and frequencies of the stars in the streams using various Galactic potentials. We also explore how the Helmi streams constrain the  flattening $q$, of the Galactic dark matter halo.}
  % results heading 
   {We find that the streams are split into substructures in integrals of motion space, most notably into two clumps in angular momentum space. The clumps have consistent metallicity distributions and stellar populations, supporting a common progeny. In all the realistic Galactic potentials explored, the Helmi streams stars  depict a diffuse distribution close to $\Omega_z/\Omega_R \sim$  0.7.
   At the same time, the reason for the substructure in angular momentum space appears to be a $\Omega_z:\Omega_\phi$ resonance close to the 1:1. This resonance is exactly the 1:1 in the case that the (density) flattening of the dark halo is $q=1.2$. For this halo shape the substructure in angular momenta is also long lasting. }
  % conclusions heading  
   {Our findings suggest that the structure of the Galactic potential leaves a clear imprint on the properties of phase-mixed debris streams.}


   \maketitle
%
%-------------------------------------------------------------------

\section{Introduction}
    
Thanks to \textit{Gaia} data \citep{gaia-dr2}, the debris of several
mergers has recently been identified in the halo near the Sun: the dominant
\textit{Gaia}-Enceladus \citep[][]{belokurov2018,koppelman2018,helmi2018},
the Helmi streams \citep{helmi1999b,koppelman2019}, Sequoia
\citep{myeong2019}, and Thamnos \citep{koppelman2019b}.  Farther out
in the halo, the proposed debris of several smaller mergers has also
been identified: Wukong, Aleph, Arjuna and I’itoi \citep[see][although the latter are
plausibly related to \textit{Gaia}-Enceladus, see \citealt{Naidu2021}]{naidu2020}.
The basic concept behind the identification of phase-mixed debris from
accreted galaxies is that they are apparent as clumps in the integrals
of motion space, such as energy and angular momenta, even long after the
merger is completed \citep{helmi2000}.

Not all substructure however has an accreted origin.
The Sagittarius dwarf galaxy seems to be the cause of a large amount of substructure, such as the phase spiral \citep{antoja2018}, ridges in $V_\phi$-$R$ space \citep{ramos2018,laporte2019} and several low-latitude over-densities \citep[e.g. the Monoceros Ring,][]{laporte2018}. Meanwhile, the rotating bar at the centre of our galaxy is, for example, the cause of the Hercules stream in the nearby disc \citep{dehnen2000} and also has an impact on the morphology of the stream originating in the halo globular cluster Pal 5 \citep{pearson2017}.

The latter substructures are induced by time dependent effects that cause changes in a star's orbit and e.g. in the case of the bar, forces 
stars on specific orbital families. The underlying (static) Galactic potential can 
also cause ``observable'' substructure in the space of orbital parameters.
For example, \citet{amarante2020} and \citet{koppelman2021} have shown that
the wedges seen in $z_{\textrm{max}}$-$r_{\textrm{max}}$ space \citep[see e.g.][]{haywood2018} are related to the transitions between different orbital families associated to 
resonances and chaotic regions in our galaxy. 

How resonances and orbital families are populated in a system can in principle be used to constrain the form of its gravitational potential, for example, the shape of its halo \citep{valluri2012}. Furthermore, streams near a resonance may remain spatially coherent on longer timescales because their stars spread out more slowly  \citep{vogelsberger2008}.
However, if the tidal debris is initially at the boundary between two resonances (a separatrix) then, depending on the size of the system, the stars may evolve on widely different orbits and diverge much faster \citep{price-whelan2016,mestre2020,yavetz2021}, making such streams more difficult to detect.
Understanding these different effects and mapping out the orbital resonances in the Galactic potential is thus important for the identification and interpretation of substructures.

In this paper we revisit the Helmi streams debris with the recently
released \textit{Gaia} EDR3 data. First identified by
\citet[][see also \citealt{chiba2000,kepley2007,smith2009}]{helmi1999b} 
and characterised further with \textit{Gaia} DR2 by
\citet[][hereafter K19]{koppelman2019}, the Helmi streams are thought
to be the debris of a massive dwarf galaxy
M$_{*} \sim10^8$~M$_{\odot}$ that was accreted $\sim$ 5-8 Gyrs
ago. Here we show that the stars' distribution in angular momentum
space exhibits substructure and we investigate its origin. We
demonstrate that the debris stellar populations in the substructures are indistinguishable from one another, and put forward the case that the substructure results from the presence of orbital resonances associated to the Galactic potential, 
particularly  $\Omega_z/\Omega_\phi \sim 1$.
We find in fact, that if the flattening of the dark matter halo component in the \citet{mcmillan2017} Milky Way mass model is $q = 1.2$ (in the density), this ratio is exactly 1. We thus argue that this is likely the preferred shape of the halo for the region probed by the Helmi streams, namely within $\sim$ 5 - 20 kpc from the Galactic centre.

\begin{figure}
    \centering
    \includegraphics[width=0.8\linewidth]{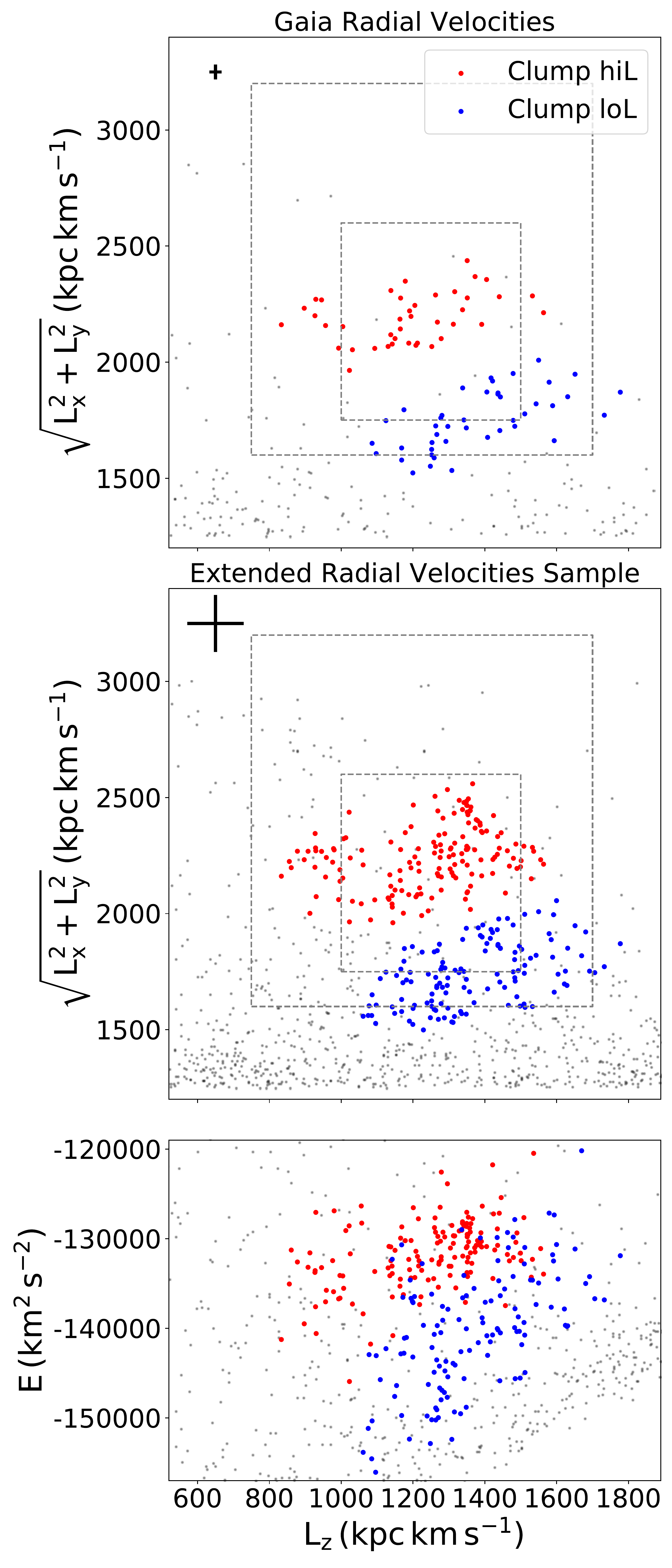}
    \caption{Helmi stream stars within 2.5 kpc are distributed in $L_z-L_\perp$ in two clumps: hiL (red) and loL (blue). The remainder of the stars in our halo sample in the 2.5 kpc volume are shown in black. Dashed lines show the selections used previously by K19 for the Helmi streams debris. The top panel contains only the stars with radial velocities from \textit{Gaia} while the middle panel contains all of the stars in the sample. The average uncertainties of the \textit{Gaia} RVS stars (27.7\% of the extended sample) and the LAMOST LRS stars (60\% of the extended sample) are shown by the black error bars in the top and middle panel respectively. The uncertainties of the remaining stars in the sample (12.3\%) are similar to those of the RVS.
    The bottom panel shows how the stars are distributed in $L_z$-$E$ space.}
    \label{fig:Lz_Lperp_Selection}
\end{figure}

This paper is structured as follows, Sect. \ref{sec:data} outlines the data used to select the Helmi streams, combining \textit{Gaia} EDR3 with spectroscopic surveys, and in Sect.~\ref{sec:results} we present our results. Specifically, in Section \ref{sec:common_origin} we show that all of the substructures in angular momentum space have consistent stellar populations in agreement that all the stars are Helmi streams debris. We explore the origin of the substructure in Sect. \ref{sec:frequency} using orbital frequency analysis (which is described in detail in Appendix~\ref{sec:appendix}), while in Section \ref{sec:potential} we demonstrate the sensitivity of the substructure to the flattening of the Galactic dark matter halo. Then in Sect. \ref{sec:conclusion} we present our conclusions.

%--------------------------------------------------------------------
\section{The data}\label{sec:data}

\subsection{Generalities}
The latest data release of \textit{Gaia}, EDR3, has provided more
accurate proper motions and parallaxes and hence more accurate
six-dimensional (6D) phase-space information for nearby stars
\citep{GaiaCollab2020a}. This data-set contains $\sim$ 1.7 billion
sources of which 7,209,831 stars have the full 6D information
(position on the sky, proper motions, parallax and radial velocity
from DR2).  Selecting objects with \verb|parallax_over_error > 5| and \verb|RUWE < 1.4| 
leaves %6,061,394 stars in total 
5,709,139 stars within a 5 kpc volume around the Sun (and 4,496,187 within 2.5 kpc). We extend this sample with radial velocities from spatial
cross-matches with the following spectroscopic surveys: APOGEE DR16
\citep{ahumada2020}, LAMOST DR6 \citep{cui2012}, Galah DR3
\citep{Buder2020} and RAVE DR6 \citep{steinmetz2020}. In total, this
results in %9,563,250 stars in total 
9,148,793 stars in the 5 kpc volume (respectively 7,531,934 within 2.5 kpc) with 6D information and 
high quality parallaxes. We use where possible the \textit{Gaia} measured radial
velocities, supplemented then by Galah, APOGEE, RAVE and finally
LAMOST\footnote{The LAMOST LRS radial velocities have
been corrected for the offset of 7.9 km\,s$^{-1}$ with respect to the
other surveys, see \url{http://dr6.lamost.org/v2/doc/release-note}} in that order of preference. 
These spectroscopic surveys also provide metallicities [Fe/H] for some of
the stars in this sample and we use the LAMOST low resolution (LRS)
metallicities for the analyses presented in this paper (because of the large numbers of targets and to ensure a uniform metallicity scale).

\begin{figure*}
    \centering
    \includegraphics[width=0.85\linewidth]{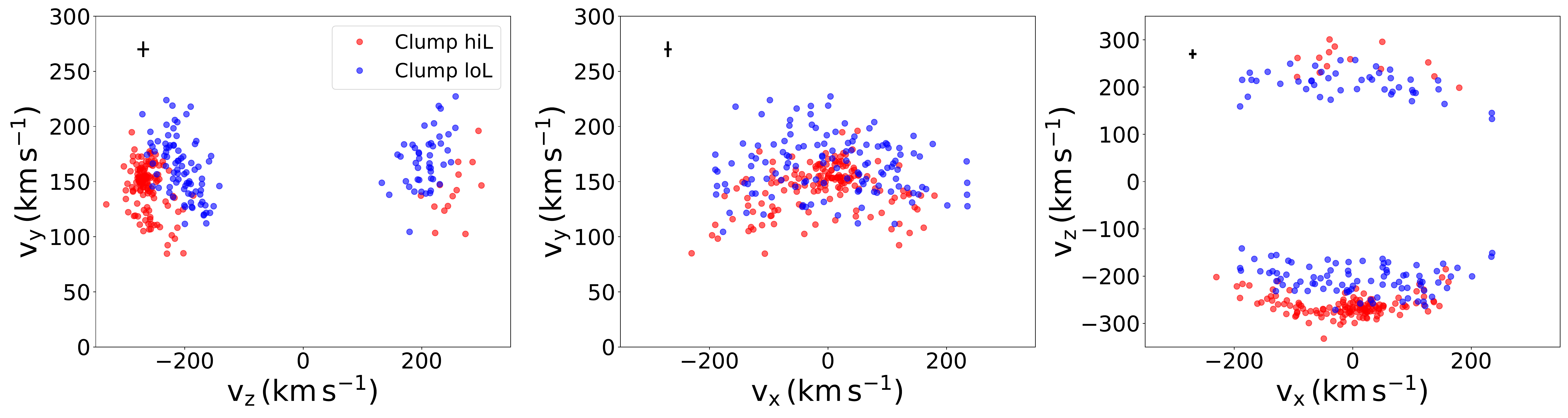}
    \caption{Distribution in velocity space of the stars in the clump-hiL (red) and clump-loL (blue) associated with the Helmi streams and located within 2.5 kpc. Both clumps populate the negative and positive $v_z$ streams characteristic of the Helmi streams debris. The average velocity uncertainties of these stars are shown by the black error bars in the top left corner of each panel.}
    \label{fig:velocities}
\end{figure*}

The quality cut on the parallaxes allows us to use the inversion of the parallax to estimate the distance, after applying a zero point offset of $-17\mu$as \citep{Lindegren2020}. We correct the stars for the solar motion using 
$ (U, V, W )_\odot$ = (11.1, 12.24, 7.25) km\,s$^{-1}$ \citep{schonrich2010}
and for the motion of the local standard of rest (LSR) using $v_{\textrm{LSR}}$ = 232.8\,km\,s$^{-1}$ \citep{mcmillan2017}.
Both the Galactocentric Cartesian and cylindrical positions and velocities of the stars are calculated assuming $R_{\odot}$ = 8.2\,kpc \citep{mcmillan2017} and $z_{\odot}$ = 0.014\,kpc \citep{binney1997}.
We define the coordinate system such that $x$ point towards the Galactic centre, $y$ in the direction of motion of the disc and positive (negative) $z$ is the height above (below) the disc. 
Angular momenta, $L_z$ and $L_{\perp}$ = $\sqrt{L_x^2 + L_y^2}$\,, are calculated for the stars with the sign of $L_z$ flipped such that it is positive for prograde orbits. Energy, $E$, is computed with \texttt{AGAMA} \citep{vasiliev2019} and using the \citet{mcmillan2017} potential for the Milky Way. This potential is axi-symmetric and made up of a stellar thin and thick disc, HI gas disc, molecular gas disc, a bulge and an NFW halo, which by default has a  spherical shape.

\subsection{Helmi Streams Selection}
Following K19 we select the Helmi streams debris in
$L_z$ vs $L_{\perp}$ space, and apply a cut in energy 
$E < -1.2 \times 10^5 $~km$^2$\,s$^{-2}$. Fig.~\ref{fig:Lz_Lperp_Selection} shows the stars
selected (in colour) in this space compared to the entire sample
(black) with the dashed lines outlining the selections made in K19. By
eye (and also using a clustering algorithm, L\"ovdal et al.~2021, subm)
it is clear that the debris separates out into two clumps, one with
high $L_{\perp}$ which we name clump-hiL (red), and the lower $L_{\perp}$ that we
refer to as clump-loL (blue). This figure includes all stars within a
2.5~kpc volume around the Sun but the substructure is also present for
stars with distances up to 5~kpc. The two
clumps and the gap are seen clearest in the top panel of
Fig.~\ref{fig:Lz_Lperp_Selection} which includes stars with radial
velocities from \textit{Gaia} only, and whose average radial velocity
uncertainty is 2.1 km\,s$^{-1}$. The few stars in the middle panel of
Fig.~\ref{fig:Lz_Lperp_Selection} that seem to populate the gap
region, $L_{\perp} \sim 1800 - 2000$~kpc\,km\,s$^{-1}$, all have
LAMOST LRS radial velocities and much larger average uncertainties of
12.7~km\,s$^{-1}$. The bottom panel of Fig.~\ref{fig:Lz_Lperp_Selection} shows that the two clumps overlap to some extent in $L_z$-$E$ space.

We now proceed to associate stars to each of the clumps as follows. Clump-hiL 
stars are selected using an ellipse centered on 
($L_z$, $L_{\perp}$) = (1225,2255) kpc\,km\,s$^{-1}$ with major/minor axis lengths of 855 and 570 kpc\,km\,s$^{-1}$ respectively, rotated by an angle of 30 degrees anti-clockwise. Clump-loL stars are selected using an ellipse centered on ($L_z$, $L_{\perp}$) = (1420,1780) kpc\,km\,s$^{-1}$ with major/minor axis lengths of 860 and 
430~kpc\,km\,s$^{-1}$ respectively, also rotated by an angle of 30 degrees anti-clockwise. There are 154 stars in clump-hiL and 130 stars in clump-loL within the 2.5 kpc volume. 
Although the remainder of the analysis presented is for this volume, we check the results remain valid for a larger volume of 5 kpc. 

\begin{figure}[ht]
    \centering
    \includegraphics[width=0.85\linewidth]{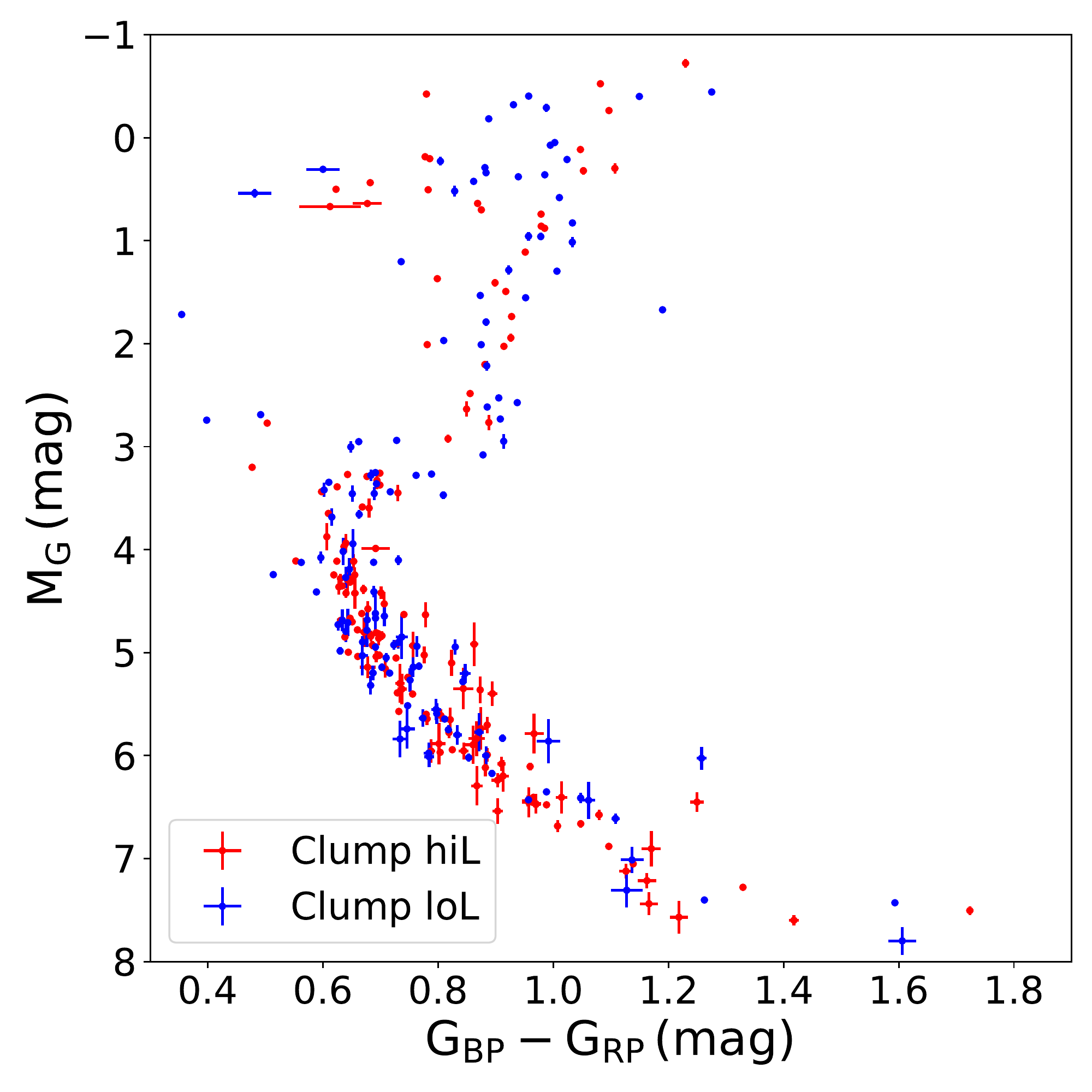}
    \caption{Colour-absolute magnitude diagram of Helmi stream members within 2.5 kpc. Clump-hiL stars are shown in red and clump-loL in blue.}
    \label{fig:CMD}
\end{figure}

\begin{figure}[ht]
    \centering
    \includegraphics[width=0.85\linewidth]{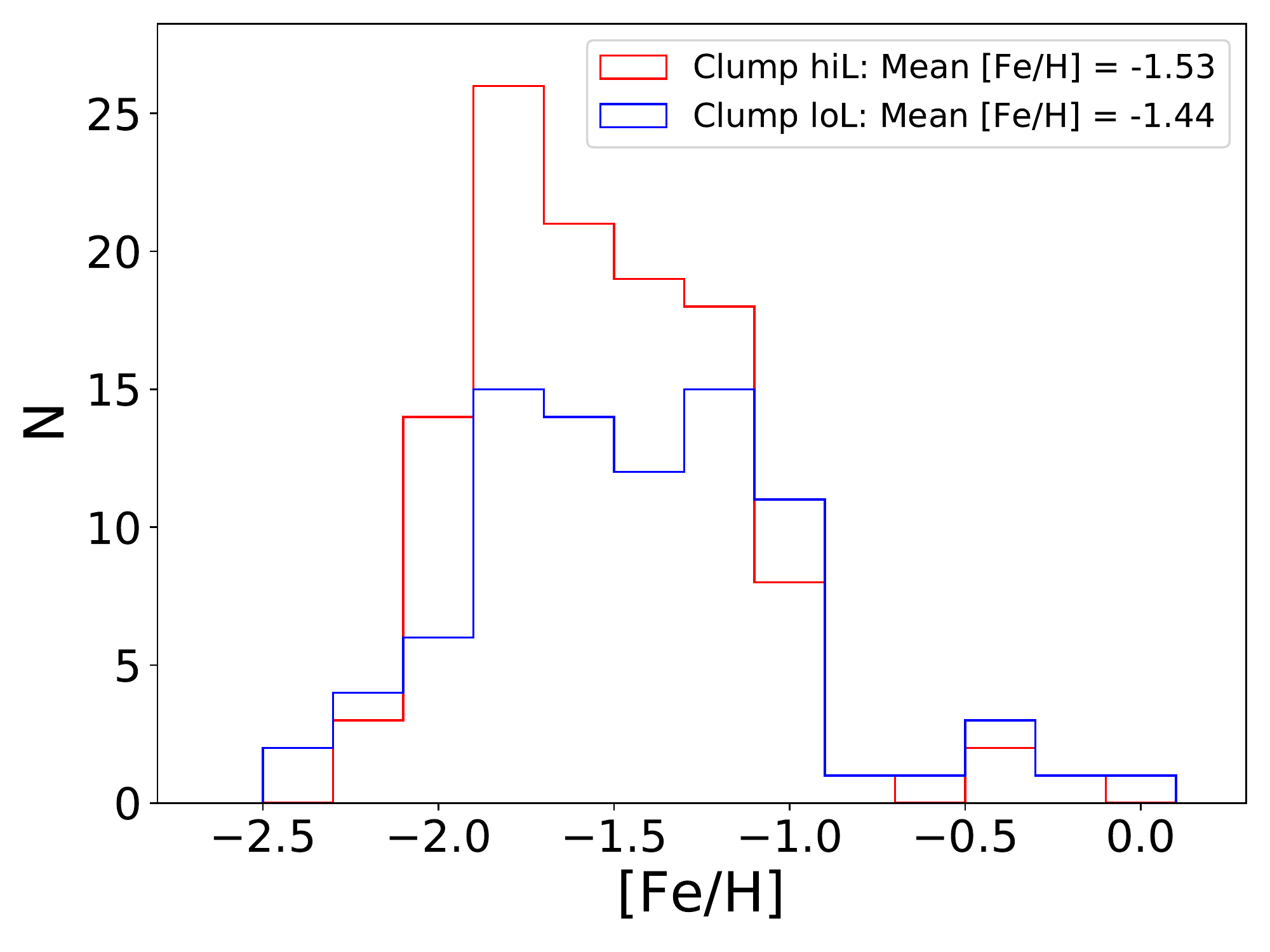}
    \caption{Metallicity distributions for the stars in the Helmi streams in clump-hiL (red) and clump-loL (blue), where available from LAMOST LRS. Both clumps exhibit similar distributions showing a broad range in metallicities peaking around [Fe/H] $\sim-1.5$.}
    \label{fig:MDF}
\end{figure}

\begin{figure*}
    \centering
    \includegraphics[width=0.9\linewidth]{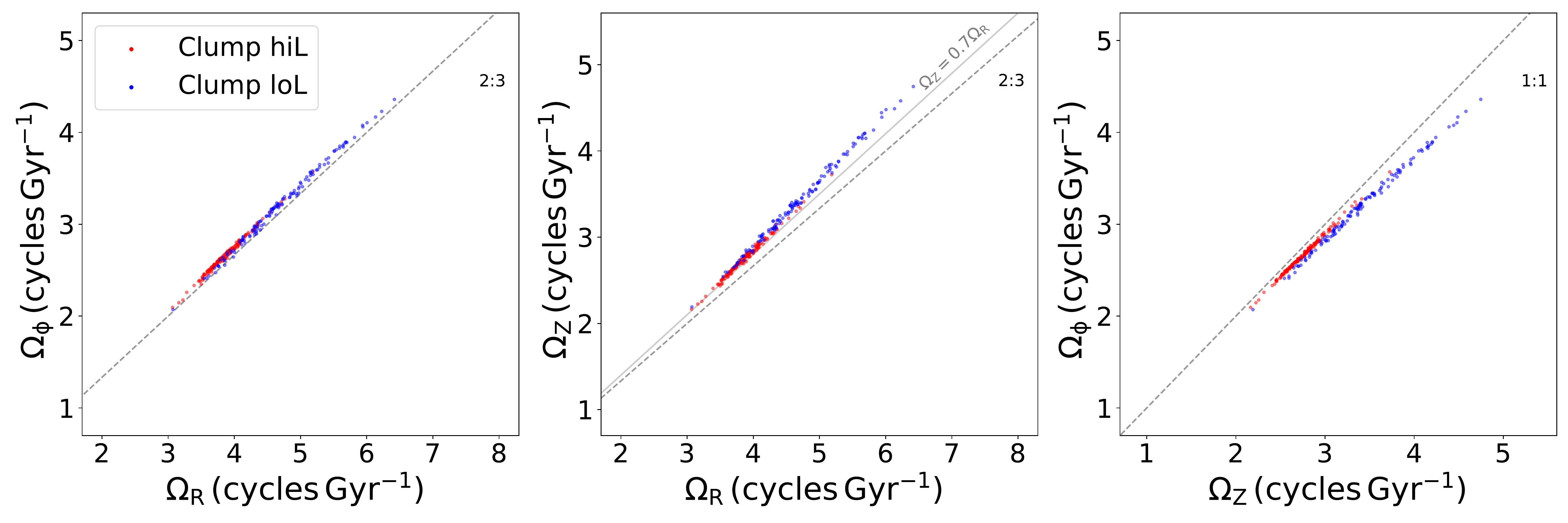}
    \caption{Frequency space for the two angular momenta clumps. Red is clump-hiL, higher L$_\perp$, and blue clump-loL. The dashed lines show the relevant resonances that the Helmi streams populate or come close to.}
    \label{fig:Freqeuncies}
\end{figure*}

Fig.~\ref{fig:velocities} shows the Cartesian velocities of the stars
colour-coded according to their membership to the clumps. Their
velocity distributions are consistent with those previously reported
for the Helmi streams. We observe negative and positive $v_z$ kinematic
groups that are populated in similar numbers by stars from the hiL
and loL clumps. In fact the asymmetry in the number of stars
(previously used by K19 to estimate the time of accretion) remains, also when considering the two
clumps in angular momentum. Note that stars with lower $L_\perp$
typically have lower $|v_z|$, since $L_\perp \sim R |v_z|$ at the
location of the Sun. We also note that some of the stars in the hiL
clump with negative $v_z$ appear to be in a kinematically cold subclump.
This tight group of clump-hiL stars likely corresponds to the S2 stream reported by \citet{myeong2018}, and can be seen to be simply part of the Helmi streams.

%--------------------------------------------------------------------
\section{Results}\label{sec:results}

\subsection{Common Origin}\label{sec:common_origin}
We now investigate the stellar populations and metallicity distributions of the 
two clumps in angular momentum space to establish if 
they have a common origin.  Fig.~\ref{fig:CMD}
shows the colour-absolute magnitude diagram (CaMD) of the two clumps,
hiL and loL in red and blue respectively.  We have propagated the errors on the fluxes and parallax into the errors on the colour (G$_{BP}$
- G$_{RP}$) and M$_G$.  The
stars that have a 6 parameter astrometric solution have had a
correction applied to the $G$ band photometry following
\citet{GaiaCollab2020a}. The observed magnitudes have also been
corrected for extinction using the 2D \citet{schlegel1998} dust map
and applying the \textit{Gaia} pass-bands extinction
coefficients\footnote{\url{http://stev.oapd.inaf.it/cgi-bin/cmd_3.4}}
\citep{cardelli1989,odonnell1994}. Fig.~\ref{fig:CMD} reveals no differences
in the CaMD of the clumps.

Both clumps also show similar metallicity distributions, as
illustrated in Fig \ref{fig:MDF}, with clump-hiL (113 stars) having a
mean [Fe/H]~$\sim-1.53$ and clump-loL (86 stars) having $\sim-1.44$, based on their LAMOST LRS [Fe/H] measurements.
A Kolmogorov–Smirnov (KS) test comparing the metallicity distributions of the two clumps yields a probability of 0.13 that the distributions are consistent with each other.

%-----------------------------------------------------------------
\subsection{Frequency Analysis}\label{sec:frequency}

The analysis of the previous section demonstrated that the stars in the two
angular momenta clumps have statistically indistinguishable stellar populations
and this supports a common origin. Thus, to explore the possible
causes of the split in angular momentum space we now investigate in more
depth the orbital properties of the two clumps. To this end we
integrate each star's orbit in the \citet{mcmillan2017} potential with
\texttt{AGAMA} \citep{vasiliev2019}. We integrate for $\sim$100
Gyr %(97.87 Gyr)
in total, outputting the positions and velocities at regular time
steps of $\sim$1 Myr. %(0.9778 Myr)
% We output 10$^5$ + 1 points along the orbit integration 

We investigate the frequencies of the orbits by using a variation on the cylindrical
polar coordinates (namely Poincar\'{e}'s symplectic polar variables)
following \citet{valluri2012} and \citet{koppelman2021}.  We input the
positions and velocities in this coordinate system as
complex time series into \texttt{SuperFreq} \citep{price2015b}, that
numerically calculates the frequencies; $\Omega_z$, $\Omega_R$,
$\Omega_\phi$.  We define 
$\Omega_\phi$ to be positive for stars that orbit in the direction of Galactic rotation.  We checked that the frequencies
remain stable when splitting the full time interval into 3 equal sub-intervals,
indicating that the orbits are regular. 

\begin{figure}[ht]
    \centering
    \includegraphics[width=0.9\linewidth]{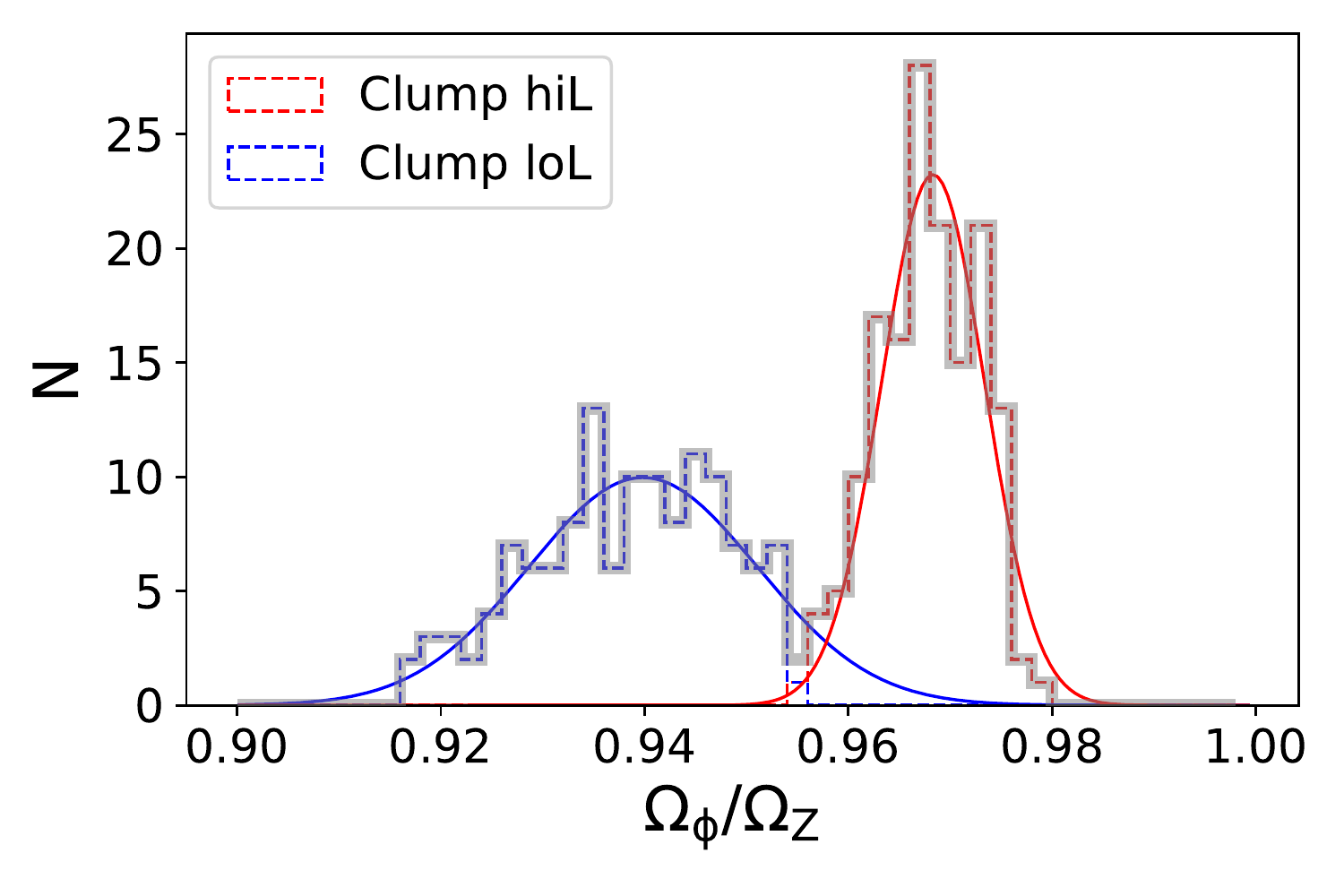}
    \caption{Distribution of $\Omega_\phi/\Omega_z$ for all stars in the Helmi streams, with clump-hiL and loL indicated in red and blue, respectively. The sum of two Gaussians fit their distributions well.  Clump-hiL stars correspond to the tight component centred on $\Omega_\phi$/$\Omega_z$ = 0.968 and clump-loL to the more diffuse component.}
    \label{fig:freq_ratio}
\end{figure}

\texttt{SuperFreq} is an implementation of the Numerical Analysis of Fundamental Frequencies algorithm \citep[originally pioneered by][]{laskar1990,laskar1993}, and determines the frequencies for the three different components $z, R, \phi$ by selecting the frequency with the highest amplitude in the Fourier transform of the time series. The resulting distributions of $\Omega_z$ and $\Omega_\phi$ for the Helmi streams stars are well-behaved, being relatively smooth and continuous as expected. However, the distribution in $\Omega_R$ depicts two branches (see Fig. \ref{fig:fR_E_two_branhces}), one of which (populated by 38\% of the stars) would seem to correspond to a $\Omega_z:\Omega_R$ = 1:2 resonance. 
If we use an alternative implementation of the NAFF algorithm by \citet{valluri1998} \citep[see also][]{valluri2010,valluri2012}, simply called \texttt{NAFF}, however, only 10\% of the stars are found to be located on the branch associated to this resonance.
Inspection of the $R(t)$ for stars on and off the resonance do not  warrant a significantly different $\Omega_R$, nor do we see evidence that there is a resonance between $R$ and $z$. Furthermore, if we proceed to action-angle space, and determine the angular frequencies using the St\"{a}ckel fudge approximation implemented in \texttt{AGAMA}, we also find a single branch and a continuous distribution of $\Omega_R$ for all the stars in the Helmi streams (see Appendix \ref{sec:appendix} for a detailed discussion). Since the frequency analysis performed by \texttt{SuperFreq} does identify what we would consider the true $\Omega_R$ but with slightly lower power, for the stars for which this happens, we choose to adopt the frequency in $R$ that is of the second highest amplitude, such that all of the stars on the same $\Omega_R$ branch.

Fig.~\ref{fig:Freqeuncies} shows the frequencies colour-coded by
angular momenta clump with clump-hiL in red and clump-loL in blue.
{\it All} of the stars
in clump-hiL, appear to be on a resonance close to
$\Omega_\phi$~:~$\Omega_z$  $\sim$~1:1 as can be seen in the third panel
of Fig.~\ref{fig:Freqeuncies}. The histogram of
$\Omega_\phi$/$\Omega_z$ for the entire 2.5~kpc sample plotted in
Fig.~\ref{fig:freq_ratio} clearly shows a peak corresponding to
$\Omega_\phi$~:~$\Omega_z$~=~0.968:1 with a thickness of $\sim$0.015. We fit a sum of two Gaussian distributions to the histogram of $\Omega_\phi$/$\Omega_z$ and confirm that all of the stars in
clump-hiL are located within 3 standard deviations from the mean of the component that represents this resonance. The second Gaussian component contains the clump-loL  stars which form a more diffuse cloud of stars at slightly lower $\Omega_\phi$/$\Omega_z$. We thus argue that the $\Omega_\phi$ : $\Omega_z \sim$~1:1 resonance, which is associated to  the gravitational potential,
must be the cause of the substructure in angular momentum
space.

\begin{figure}[ht]
    \centering
    \includegraphics[width=0.95\linewidth]{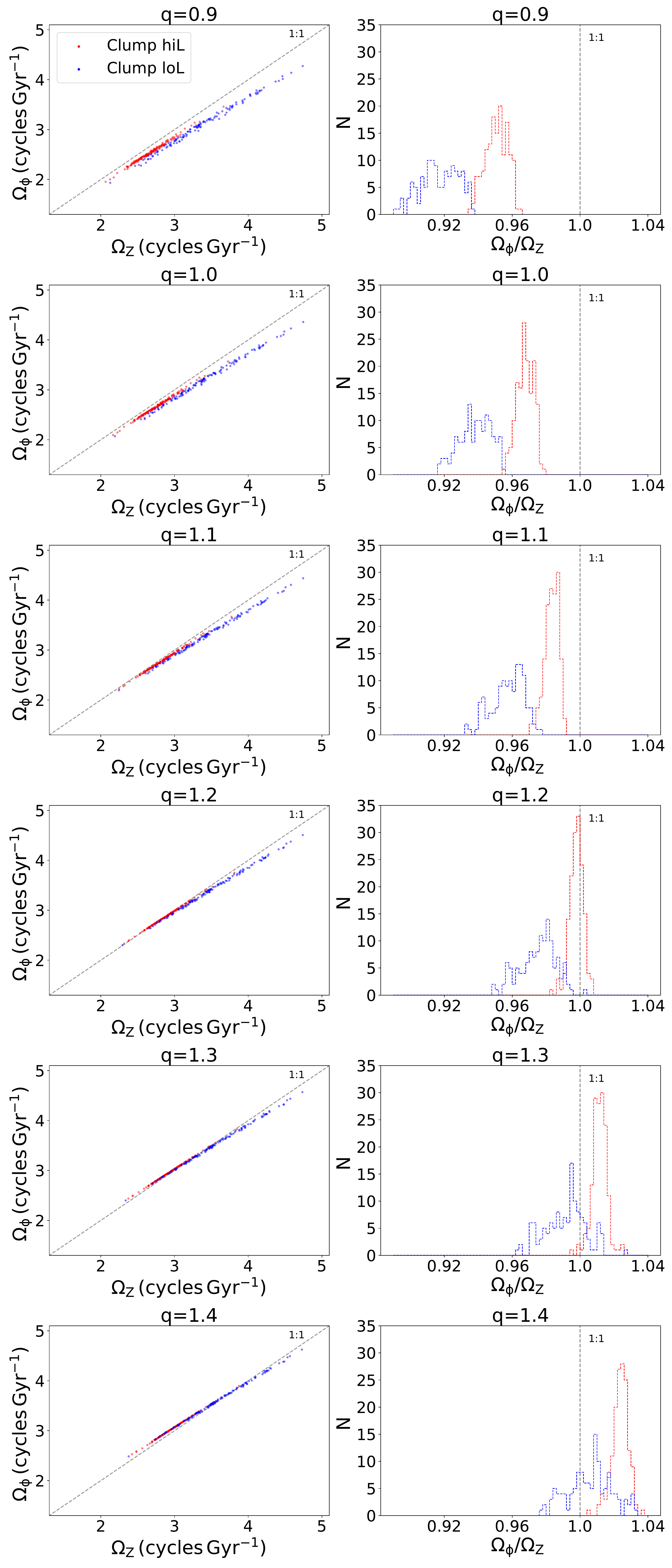}
    \caption{Frequency, $\Omega_\phi$ and $\Omega_z$, distributions (left column) and histograms of their ratios (right column) for all stars in the Helmi streams. Clump-hiL and loL stars are shown in red and blue respectively. Each row shows a different flattening added to the \citet{mcmillan2017} halo component with the axis ratio, $q$, increasing from 0.9 (oblate) to 1.4 (prolate).}
    \label{fig:freq_flattening}
\end{figure}

\subsection{A new constraint on the shape of the Galactic dark matter halo}
\label{sec:potential}

\begin{figure*}[ht]
    \centering
    \includegraphics[width=0.85\linewidth]{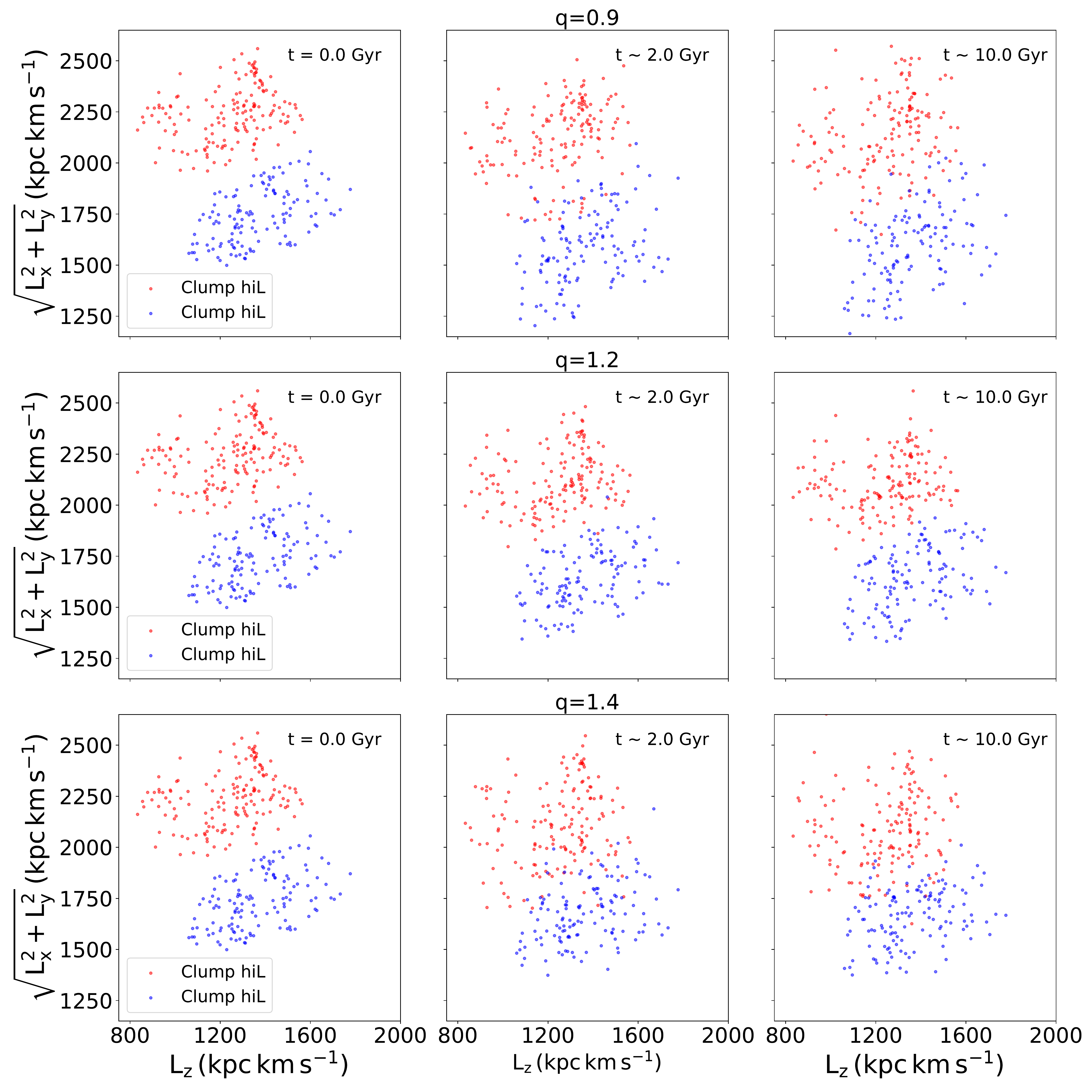}
    \caption{Evolution of $L_z-L_\perp$ over time in the \citet{mcmillan2017} potential with various flattening values $q$ for the density of the halo component. The first column shows the present day observed distribution in $L_z-L_\perp$, the second column the stars in $L_z-L_\perp$ space after $\sim$ 2 Gyr integration time and the third column after $\sim$ 10 Gyr. The first row shows the results for the model with the lowest $q$ (0.9) and the bottom row the highest $q$, with the best estimate for $q = 1.2$ shown in the middle row.}
    \label{fig:LzLperp_time}
\end{figure*}

\begin{figure*}
    \centering
    \includegraphics[width=0.85\linewidth]{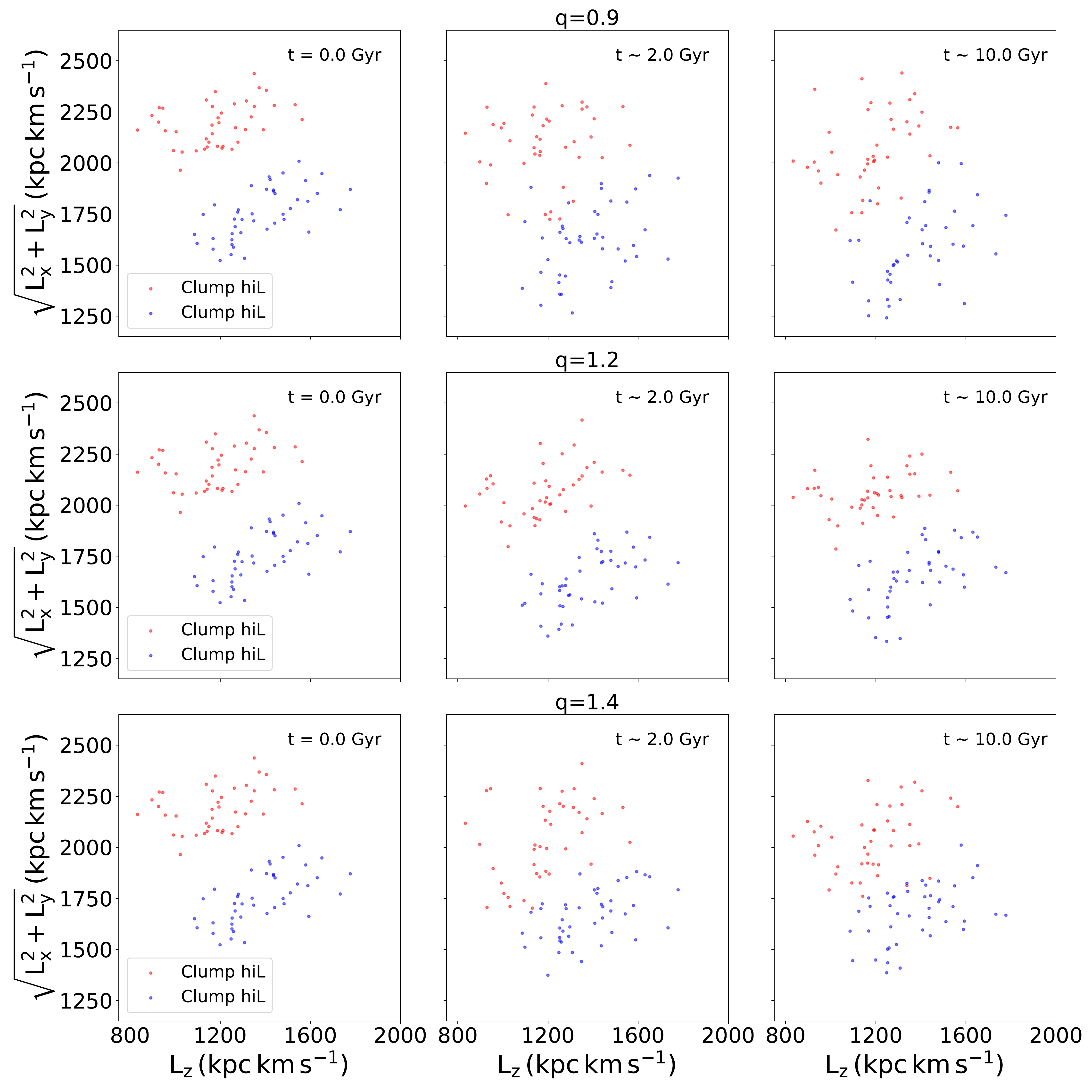}
    \caption{Same as Fig. \ref{fig:LzLperp_time} but for the Gaia radial velocity sample that have significantly lower uncertainties.}
    \label{fig:LzLperp_time_Gaia_RVS}
\end{figure*}

In the chosen \citet{mcmillan2017} potential 
the gap between 
the two clumps in angular momentum space remains visible over time, especially when considering only \textit{Gaia} radial velocity stars
where the gap is clearest. 
Although $L_\perp$ is not constant in this potential, its variation with time is not large enough to erase the gap (for over 2 Gyr) nor to move stars from one clump to the next (for at least 10 Gyr).

We checked that the $\Omega_\phi$/$\Omega_z$ resonance holds and is similarly populated when calculated with orbits integrated in various other potentials. In the Galactic potentials of \citet{bovy2015}, \citet{piffl2014} and \citet{price2017}, the clump-hiL stars correspond to the tighter group of stars that are on a $\Omega_\phi$ : $\Omega_z$ resonance close to the 1:1. 
In the static triaxial \citet{vasiliev2021} potential\footnote{We use the fiducial model with a static twisted halo and no LMC, which these authors suggest is the best static potential fitted on the Sagittarius stream.} a branch close to the above-mentioned resonance is apparent, although it is more diffuse. More importantly the gap in angular momentum space disappears after $<$100 Myr in the orbit integration. This suggests that this model provides a less accurate representation of the Galactic potential in the region probed by the orbits of the Helmi streams
stars, if we make the reasonable assumption that we are not living at a special time in Galactic history.

In order to investigate if there is any sensitivity to the shape of the Galactic halo as suggested by \citet{valluri2012}, 
 we introduce flattening into the dark matter halo component of the \citet{mcmillan2017} potential. 
The density of this component follows the NFW form: 
\begin{equation*}
\rho = \rho_o \bigg(\frac{\tilde{r}}{a}\bigg)^{-1} \Bigg[1 + \frac{\tilde{r}}{a} \Bigg]^{-2} 
\end{equation*}
with
\begin{equation*}
\tilde{r} = \sqrt{x^2 + y^2 + \bigg(\frac{z}{q}\bigg)^2}.
\end{equation*}
and $\rho_o$ = 8.53702 $\times \, 10^6$ M$_\odot\,$kpc$^{-3}$, $a$ = 19.5725 kpc. 

We vary $q$ (the density axis ratio of $z$ to $R$) from 0.9 to 1.4 in steps of 0.1.
We then integrate the orbits of all stars in these different potentials and calculate the orbital frequencies as before, see Sect. \ref{sec:frequency}. The different rows in Fig. \ref{fig:freq_flattening} show the results. As $q$ increases the resonance that the clump-hiL stars occupy moves closer to the $\Omega_\phi$~:~$\Omega_z$  = 1:1, up until $q = 1.2$ after which the stars move past the 1:1. The distribution of the clump-hiL stars in $\Omega_z$ - $\Omega_\phi$ space also gets narrower and more peaked as $q$ increases up to $q = 1.2$, as can be seen in the second column of Fig. \ref{fig:freq_flattening}.

We also find that as we reduce the flattening of the halo component ($q = 0.9$) the gap in $L_z$ and $L_\perp$ is no longer conserved and stars move from one clump to another, see the first row of Figs. \ref{fig:LzLperp_time} \& \ref{fig:LzLperp_time_Gaia_RVS}. As $q$ increases above 1.0, and the dark matter halo becomes more elongated, there is less mixing of stars and the gap is maintained better, up until $q = 1.2$, for which we see the least amount of mixing and that the gap is best conserved. This can be seen in the middle row of Figs. \ref{fig:LzLperp_time} \& \ref{fig:LzLperp_time_Gaia_RVS}. Then as we increase $q$ further ($q > 1.2$) we see the mixing of stars increase and the gap disappear, as shown in the bottom row of Figs. \ref{fig:LzLperp_time} 
\& \ref{fig:LzLperp_time_Gaia_RVS} for $q = 1.4$.

Since $L_z$ and $L_\perp$ can be computed directly from the observables without knowledge of the Galactic potential, but their evolution in time depends on its specific functional form, the presence of two clumps, or a long-lasting gap, serves to directly constrain the shape of the potential. Our analysis suggests that the dark matter halo component of the Galaxy in the region probed by the Helmi streams (within $\sim$20 kpc) has an elongated shape with a density $z/R$ axis ratio of $\sim 1.2$. The shape of the dark matter halo is thought to be oblate or spherical in the central regions \citep[e.g.][]{koposov2010} and gradually becoming more triaxial with its major axis along $z$ at larger radii \citep[e.g.][]{law2010}, possibly with a varying flattening as a function of radius \citep[e.g][]{,vera2013}. Our new constraint at an intermediate radius is in close agreement with measurements using the Galactic globular clusters as tracers that probes similar radii as the Helmi stream orbits \citep[$q$ = 1.30 $\pm$ 0.25,][]{posti2019}.

%-----------------------------------------------------------------

\section{Discussion and conclusions}\label{sec:conclusion}

We have shown that the local Helmi streams debris are divided into two
clumps in $L_z$-$L_\perp$ space, most clearly for stars within 2.5~kpc from the Sun. The stars in these clumps most likely share the same
progenitor as their stellar populations are indistinguishable. We have
determined that the origin of this substructure in angular momentum
space is related to the presence of an orbital resonance. Specifically the stars in the high $L_\perp$ clump are on a  $\Omega_\phi$~:~$\Omega_z$ resonance located very close to the ~1:1 resonance, while the remaining stars show a broader range of frequencies. In $\Omega_R$-$\Omega_z$ all of the stars are distributed close to $\Omega_z/\Omega_R \sim  0.7$.

We use the Helmi streams debris to put a constraint on the shape of the dark matter halo. Our findings suggest that the dark matter halo within $\sim$ 5--20 kpc of the Galactic center is elongated along the axis perpendicular to the disc ($z$) with a value of $q \sim 1.2$ (in the density). 
Evidence for this value of $q$ arises from that the stars in the high $L_\perp$ clump form the tightest distribution, populating the $\Omega_\phi$~:~$\Omega_z$  = 1:1 resonance, but are broader and offset from this integer ratio for other values of $q$. Furthermore, the evolution of the clumps and the gap in $L_z$ - $L_\perp$ space is longest lived and depicts the least amount of mixing for $q=1.2$.

We have presented the analysis of the streams for the stars
located within 2.5 kpc but we find similar results when we go farther
away and up to 5 kpc from the Sun. Going to larger
volumes with the Helmi streams is a possible way in which we could map
out the local potential. 
It is probably the first time that a ``direct'' observable such as $L_z$ - $L_\perp$ space has been used  specifically linked to a particular resonance in
frequency space for stars in the halo and we have demonstrated how it can be useful in constraining the potential.

The presence of such rich structure in terms of different resonant
families in a small region of phase-space (i.e.~that occupied by the Helmi
streams), may force us to re-think how we model the evolution of
streams. If stars in a stream are on a resonance they remain coherent for longer, but if they are just off a resonance and the stream stars orbit's fall either side of a resonant orbit then the stars are affected and the stream diverges very quickly \citep[e.g.][]{yavetz2021}.
When we model streams we do not take into account the effect of resonant orbits and simply assume that they are on regular orbits.

Our findings also add to the evidence that there can
be substructure in the orbital properties and notably in integrals of
motion space that is not a result of accretion but an effect of the
Galactic potential. If single progenitors are splitting up in this
manner then this further complicates the identification (and
characterisation) of their phase-mixed debris. Furthermore, frequency space
has been argued to be useful to constrain the time of accretion since
merger debris separates in a regular pattern
in this space associated to each individual stream crossing the
specific volume being considered \citep[e.g.][]{gomez2010}. Although we do see evidence of
multiple small clumps potentially associated to such individual
streams, the presence of resonances inducing further substructures
complicates the prospects of straightforwardly using this idea.

Several questions remain on the origin of these different
resonances. 
For example, is there a link to Sagittarius?
Previously,
\citet{koppelman2021} showed using also the McMillan potential, that Sagittarius' orbit falls on the
$\Omega_\phi$ : $\Omega_z$ = 1:1 resonance, suggesting that there may be some relation between the dynamics of Sgr and the orbits of the Helmi streams debris. On the other hand, streams on prograde orbits could also be affected by the Galactic bar, as demonstrated by \citet{pearson2017}.

\begin{acknowledgements}
We would like to thank the anonymous referee for their contributions to this article. We also thank Leandro Beraldo e Silva (and Monica Valluri) for the helpful discussions and guidance in using \texttt{NAFF}.
We acknowledge financial support from a Spinoza prize to AH. HHK gratefully acknowledges support from the Martin A. and Helen Chooljian Membership at the Institute for Advanced Study.
This work has made use of data from the European Space Agency (ESA) mission {\it Gaia} (\url{https://www.cosmos.esa.int/gaia}), processed by the {\it Gaia} Data Processing and Analysis Consortium (DPAC, \url{https://www.cosmos.esa.int/web/gaia/dpac/consortium}). Funding for the DPAC has been provided by national institutions, in particular the institutions participating in the {\it Gaia} Multilateral Agreement.

The analysis has benefited from the use of the following packages: vaex \citep{breddels2018}, \texttt{SuperFreq} \citep{price2017}, \texttt{NAFF} \citep{valluri1998,valluri2010,valluri2012}, \texttt{AGAMA} \citep{vasiliev2019}, numpy \citep{van2011}, matplotlib \citep{hunter2007} and jupyter notebooks \citep{kluyver2016}.

\end{acknowledgements}

\bibliographystyle{aa} 
\bibliography{main}

\begin{appendix}
\section{Comparison of Methods for Frequency Determination} \label{sec:appendix}

\begin{figure}[h]
    \centering
    \includegraphics[width=0.9\linewidth]{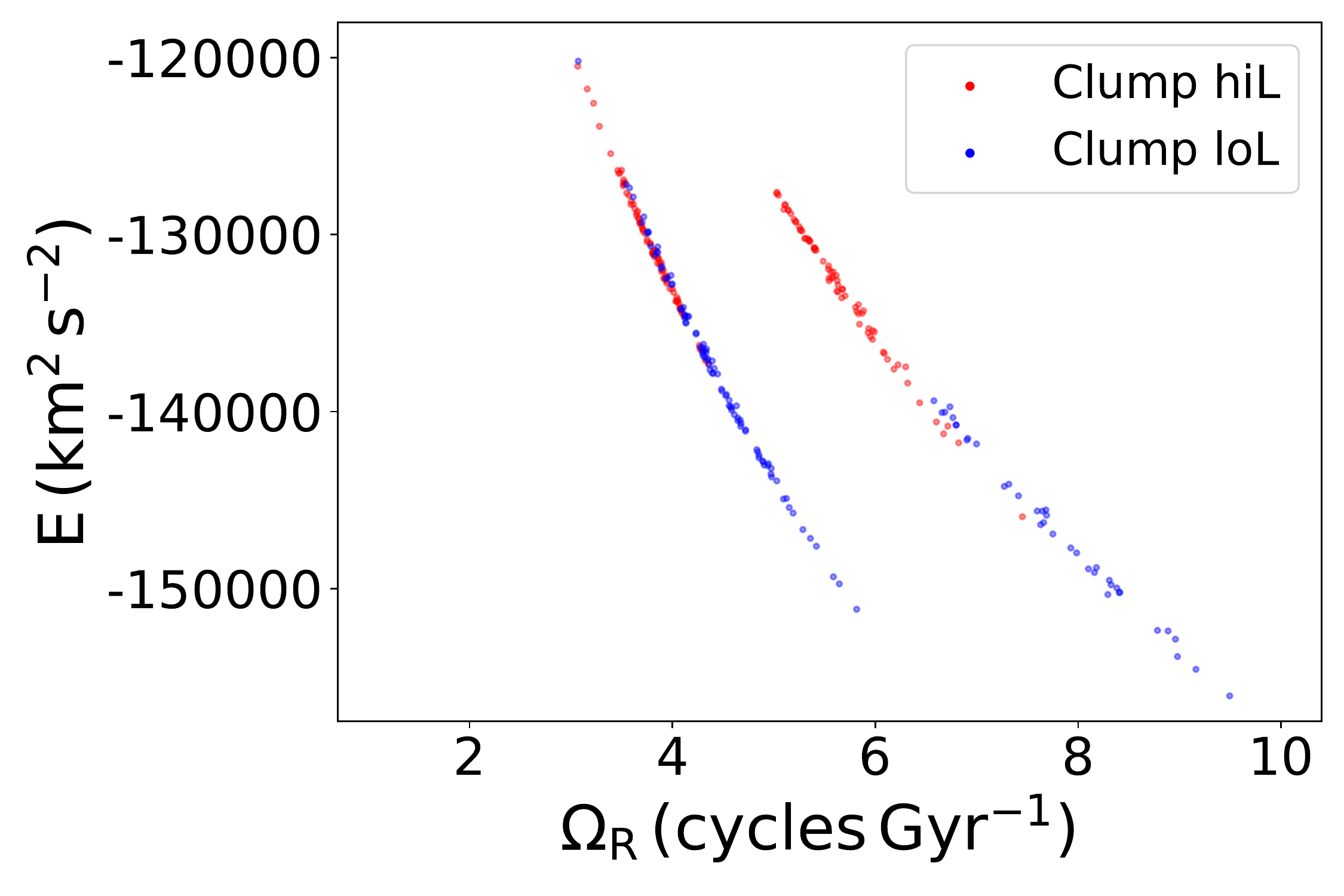}
   \caption{Distribution of radial frequencies, $\Omega_R$, from  \texttt{SuperFreq} with energy coloured according to $L_z$-$L_{\perp}$ clump. Two branches can be seen clearly. The branch at higher $\Omega_R$ corresponds to the stars that are placed onto the $\Omega_Z$ : $\Omega_R$ = 1:2 resonance.}
   \label{fig:fR_E_two_branhces}
\end{figure}

Here we compare three different methods for determining the orbital frequencies. 
Presented earlier in the paper are the frequencies determined with \texttt{SuperFreq}, \citep{price2015b} which is one implementation of the Numerical Analysis of Fundamental Frequencies algorithm. We observed that using this method the stars form two branches in $\Omega_R$ which can be seen in Fig. \ref{fig:fR_E_two_branhces} where the stars on the higher $\Omega_R$ branch are those on the $\Omega_Z$ : $\Omega_R$ = 1:2 resonance. Fig. \ref{fig:fR_E_two_branhces} shows that stars with the same energy (and very similar angular momenta) have very  different $\Omega_R$.

Another implementation of the Numerical Analysis of Fundamental Frequencies algorithm is \texttt{NAFF} by \citet{valluri1998} \citep[see also][]{valluri2010,valluri2012}. 
As already mentioned, we find differences in $\Omega_R$ for a significant fraction of the stars when comparing these two methods moving 28\% of the stars from the resonance in $\Omega_Z$ : $\Omega_R$ to the branch closer to $\Omega_Z$/$\Omega_R \sim 0.7$. For all stars, both methods show peaks in the frequency spectrum of $\Omega_R$ that would place the star on the $\Omega_Z$ : $\Omega_R$ = 1:2 resonance or on the $\Omega_Z$/$\Omega_R \sim 0.7$ branch. In some cases the amplitudes of these two peaks is very similar, possibly why there may be some discrepancy in the true $\Omega_R$.

An alternative method is to determine the frequencies ($\Omega_i$) analytically from the actions ($J_i$) using that $\Omega_i = \delta H/\delta J_i$, where $H$ is the Hamiltonian. This can be done for a St\"{a}ckel potential and so
we can approximate the \citet{mcmillan2017} to be a St\"{a}ckel potential at different points along an orbit and analytically solve for the frequencies at each point. This method is known as the St\"{a}ckel fudge method \citep{binney2012}. We use
\texttt{ActionFinder} from \texttt{AGAMA}  \citep{vasiliev2019} with the orbit inputted as $R(t)$, $z(t)$, $\phi(t)$. \texttt{ActionFinder} implements the St\"{a}ckel fudge method and gives an approximation for the actions, angles and frequencies that vary along the orbit. We then take the average over the full orbit for each star to be the frequency. The frequencies derived in this way show only one single branch in $\Omega_R$ but the same distributions in $\Omega_Z$ and $\Omega_\phi$. We check that the variations in $\Omega_R$ along the orbit are not large enough to move the stars from one branch to the $\Omega_Z$ : $\Omega_R$ = 1:2 resonance. We thus argue that the $\Omega_R$ determined by
\texttt{SuperFreq} and \texttt{NAFF} 
for the stars that are placed on the 1:2 resonance are not the true frequencies, and we take the $\Omega_R$ frequency with the second highest amplitude in the frequency spectrum, which places the stars onto a single branch.

\begin{figure}
    \centering
    \includegraphics[width=0.8\linewidth]{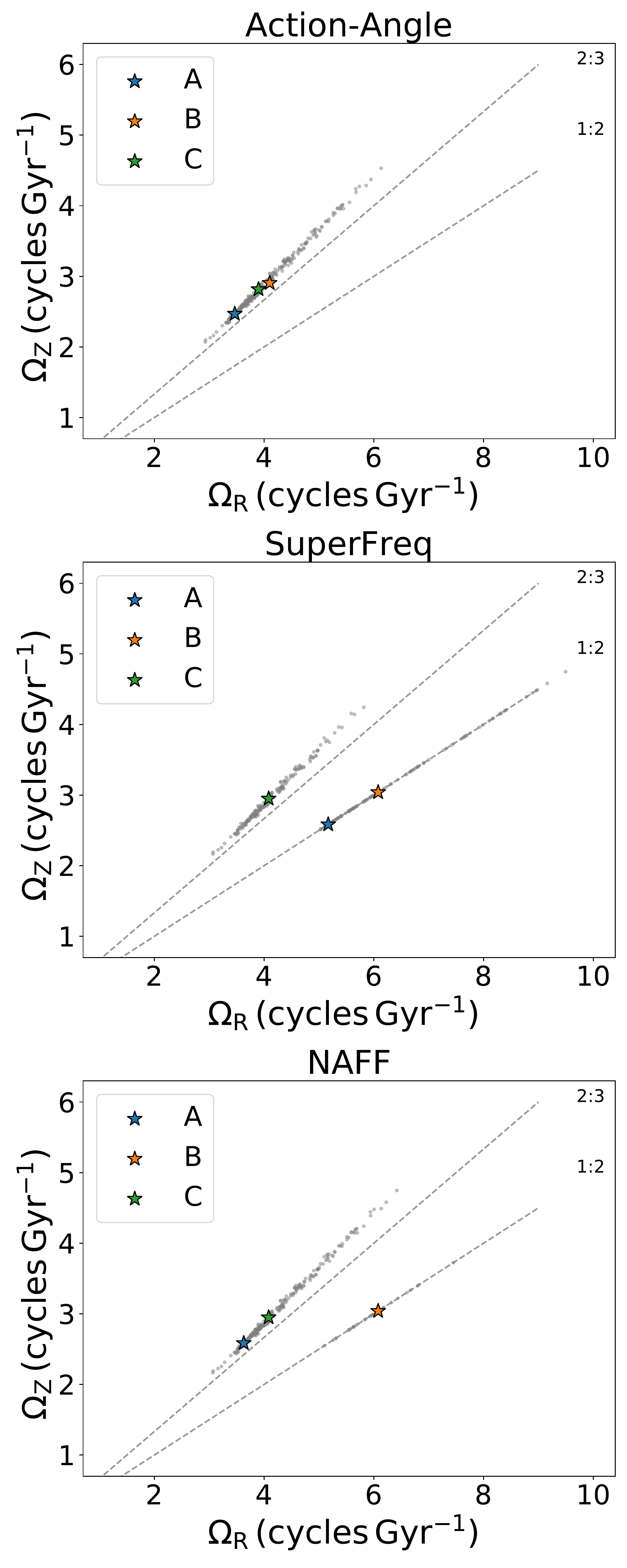}
   \caption{The first row shows the $\Omega_z$-$\Omega_R$ frequencies determined using the action-angles method in grey with three selected stars (A, B, C) in coloured star markers. These three stars are also depicted in the following two rows which show the \texttt{SuperFreq} and \texttt{NAFF} determined frequencies respectively. Star A (blue) is a star that is determined to be on the $\Omega_z/\Omega_R$ = 0.7 branch for both the action-angle frequencies and also by \texttt{NAFF}, however, \texttt{SuperFreq} disagrees placing this star on the 1:2 resonance. Star B (orange) is on the $\Omega_z/\Omega_R$ = 0.7 branch for the action-angle frequencies only, but \texttt{NAFF} and \texttt{SuperFreq} place this star on the 1:2 resonance. Finally, star C (green) is on the $\Omega_z/\Omega_R$ = 0.7 branch in all three methods.}
   \label{fig:compare_method_freq}
\end{figure}

\begin{figure}
    \centering
    \includegraphics[width=0.85\linewidth]{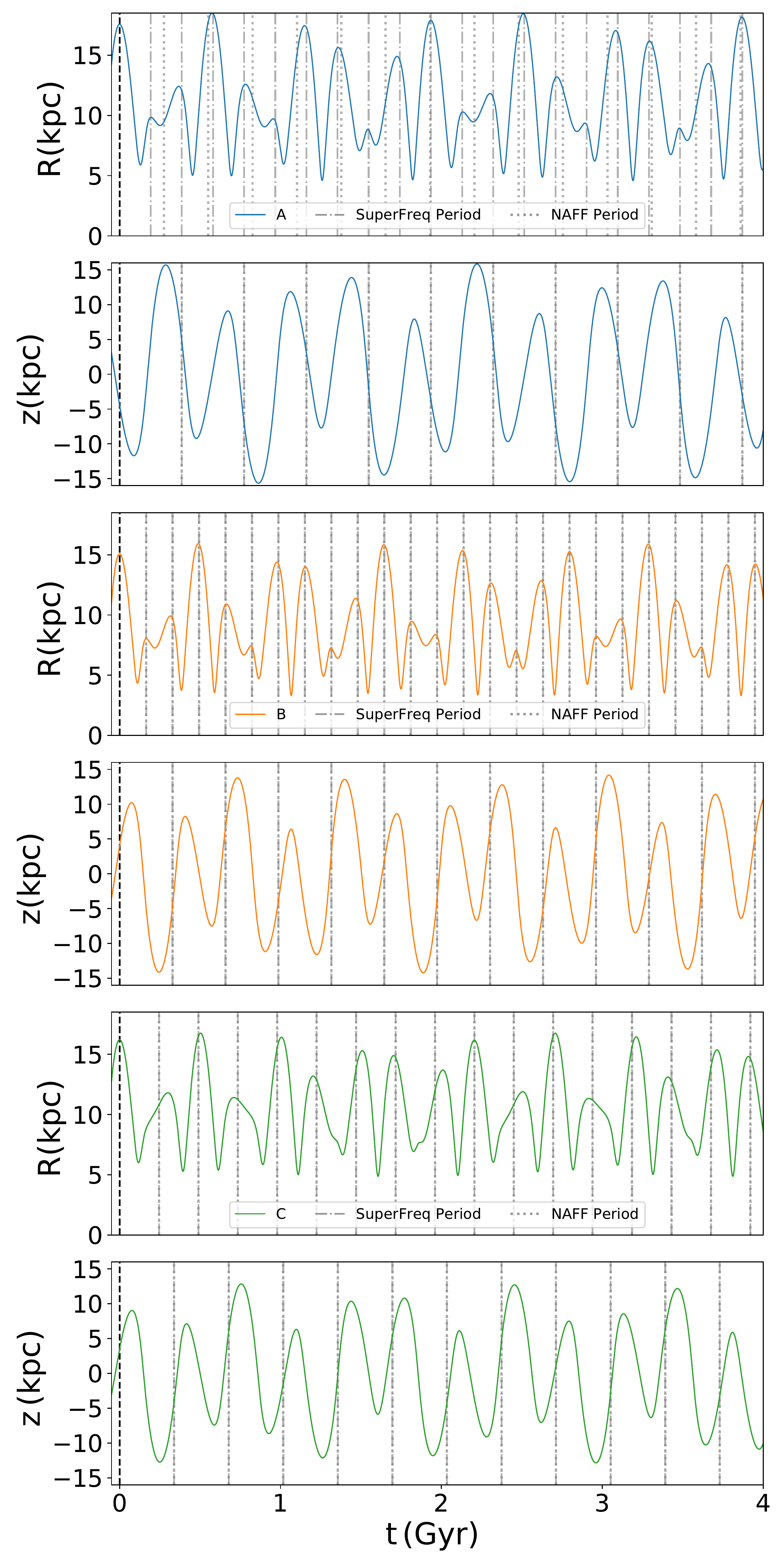}
   \caption{Evolution of $R$ and $z$ with time for the orbit integration's of the three selected stars; A, B and C. Each star's $R(t)$ is shown over a period 4 Gyr with $t$ = 0 Gyr set to the first maxima such that the three stars are in phase. The $z(t)$ is shown over the same time period for the corresponding star. Grey dash-dot and dotted lines correspond to the periods according to the frequencies derived by \texttt{SuperFreq} and \texttt{NAFF} respectively.}
   \label{fig:compare_R_z}
\end{figure}

\begin{figure}
    \centering
    \includegraphics[width=0.85\linewidth]{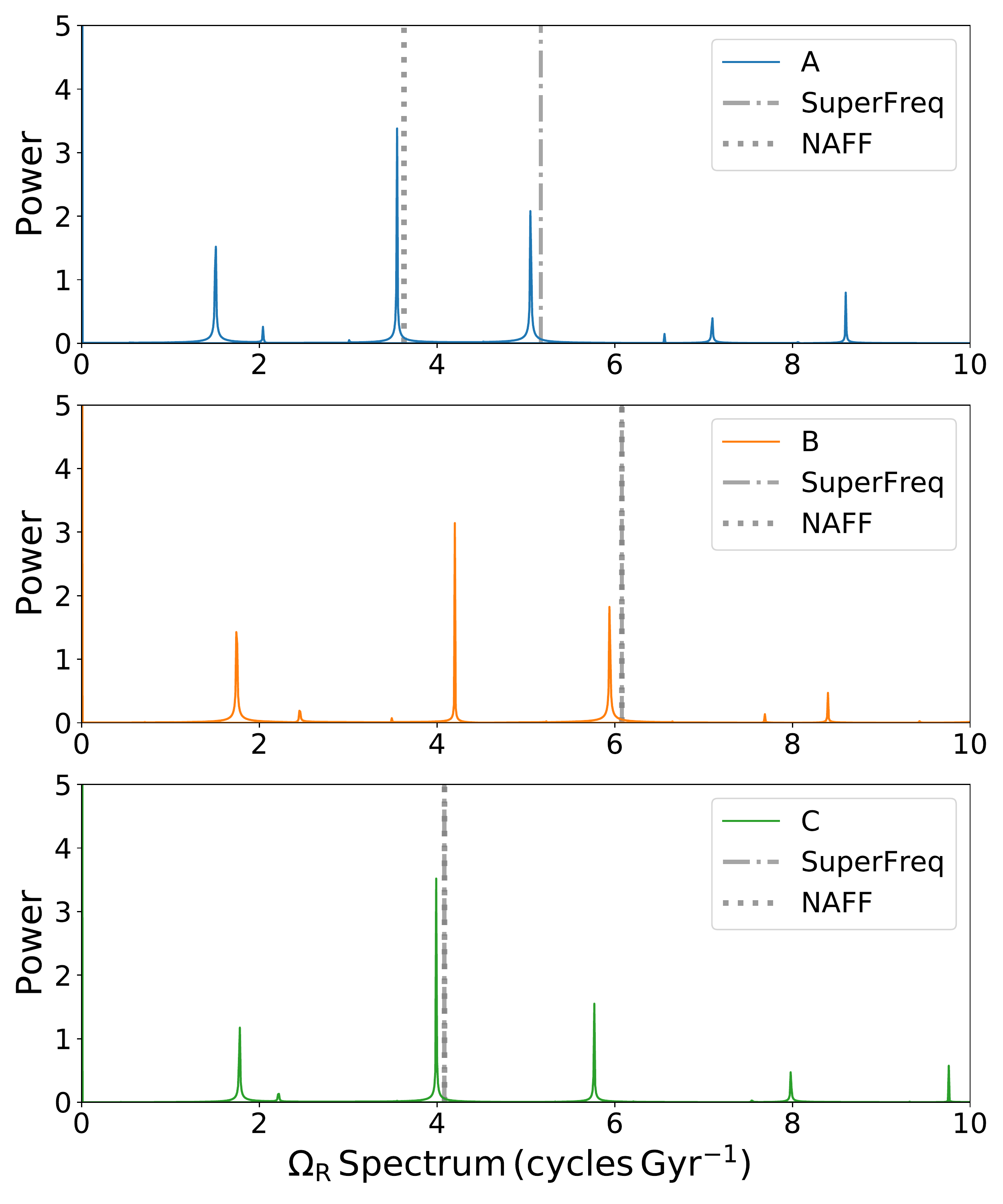}
   \caption{Fast Fourier transform of $R(t)$ for the full $\sim100$Gyr orbit integration of the three selected stars; A, B and C. Grey dash-dot and dotted lines correspond to the frequencies derived by \texttt{SuperFreq} and \texttt{NAFF} respectively. All three stars show a peak at $\Omega_R$ $\sim$4 that would place the star on the single $\Omega_Z$-$\Omega_R$ branch and another peak at a slightly higher frequency that places the star on the 1:2 resonance. }
   \label{fig:freq_spectrum}
\end{figure}

We demonstrate further the differences in the $\Omega_R$'s determined by these methods using three example stars; A, B and C. Fig. \ref{fig:compare_method_freq} shows these three stars in $\Omega_R$-$\Omega_Z$ space. Star A (blue) corresponds to a star that is on the 1:2 resonance in  \texttt{SuperFreq} but the single branch with both
\texttt{NAFF} and the action-angle frequencies. Star B (orange) is a star that is on the 1:2 resonance with both \texttt{SuperFreq} and \texttt{NAFF}, but the single branch with the action-angles. Finally, star C (green) is a star that is on the single branch in $\Omega_R$ for all three methods.

In Fig. \ref{fig:compare_R_z} we show the evolution of $R$ and $z$ with time for the orbit integration's of these three stars. We show the stars over a time period of 4 Gyr with $t$ = 0 Gyr set to the first peak such that the three stars are in phase to aid comparison. Dashed and dotted lines correspond to the periods according to the frequencies derived by
\texttt{SuperFreq} and \texttt{NAFF}. For both star B and C the two methods agree and so these lines overlap, but, for star A the periods do not line up, but we do see them overlap as the overall pattern in $R(t)$ repeats ($t$ $\sim$ 2 Gyr). Comparing the three stars $R(t)$ shown in Fig. \ref{fig:compare_R_z}, it is not clear that the stars should have as different $\Omega_R$ as suggested by the frequency identification methods.

Fig. \ref{fig:freq_spectrum} shows the Fast Fourier transform (FFT) of $R(t)$ for the three stars and the frequencies determined by
\texttt{SuperFreq} and \texttt{NAFF} shown by the dashed and dotted lines. There is not a clear difference between these three frequency spectrum's or an indication as to why a method prefers the higher frequency in some cases, placing it on the 1:2 resonance. The frequency spectrum shown in Fig. \ref{fig:freq_spectrum} is derived from $R(t)$ only, where
\texttt{SuperFreq} and \texttt{NAFF} take the complex time series of $R(t) + i\,V_R(t)$. We use Fig. \ref{fig:freq_spectrum} simply to illustrate the two peaks and show that there are no differences in $R(t)$ to suggest the difference in $\Omega_R$. If we instead look at the 
\texttt{SuperFreq} frequency spectrum then the main differences that we observed are that when a star is assigned an $\Omega_R$ that places it onto the 1:2 resonance then this is the peak with the highest amplitude e.g. stars A and B.

We also note that star's B and C have very similar energy and so it does not make sense for the two stars to have a big difference in $\Omega_R$ like  \texttt{SuperFreq} and \texttt{NAFF} suggest.

\end{appendix}

\end{document}